\documentclass[preprint]{aastex63}

\begin{document}

\title{Scattering of Giant Planets and Implications for the Origin of the Hierarchical and Eccentric Two-planet System GJ 1148}

\author[0000-0002-3237-8409]{Longhui Yuan}
\affiliation{Department of Earth Sciences, The University of Hong Kong, Pokfulam Road, Hong Kong}
\affiliation{Institut für Geophysik, ETH Zürich, Sonneggstrasse 5,
8092 Zürich, Switzerland}

\author[0000-0003-1930-5683]{Man Hoi Lee}
\affiliation{Department of Earth Sciences, The University of Hong Kong, Pokfulam Road, Hong Kong}
\affiliation{Department of Physics, The University of Hong Kong, Pokfulam Road, Hong Kong}

\begin{abstract}

The GJ 1148 system has two Saturn-mass planets orbiting around an M dwarf star on hierarchical and eccentric orbits, with orbital period ratio of $13$ and eccentricities of both planets of $0.375$.
The inner planet is in the regime of eccentric warm Jupiters.
We perform numerical experiments to study the planet-planet scattering scenario for the origin of this orbital architecture.
We consider a third planet of $0.1 M_{\text{J}}$ (Jupiter’s mass) in the initial GJ 1148 system with initial orbital separations of 3.5, 4, and 4.5 mutual Hill radii and initial semimajor axis of the innermost planet in the range of $0.10$--$0.50\,$au.
The majority of scattering results in planet-planet collisions, followed by planet ejections, and planet-star close approaches.
Among them, only planet ejections produce eccentric and widely separated two-planet systems, with some having similar orbital properties to the GJ 1148 system.
We also examine the effects of general relativistic apsidal precession and a higher mass of $0.227 M_{\text{J}}$ for the third planet.
The simulation results suggest that the GJ 1148 system may have lost a giant planet.
We also perform simulations of the general problem of the origin of warm Jupiters by planet-planet scattering.
As in the GJ 1148 simulations, a nontrivial number of stable two-planet systems are produced by ejection, which disagrees with the result from a previous study showing that two-planet systems arise exclusively through planet-planet collisions.

\end{abstract}

\section{Introduction} \label{sec:intro}

M dwarf stars have been a primary target in the search for exoplanets in recent years due to the large number of them in the galaxy. Their lower masses and luminosities compared to Sun-like stars make the planets in the so-called habitable zone (HZ) easier to detect with larger radial-velocity (RV) signal and shorter orbital period \citep{trifonov2020carmenes,Sabotta2021A&A...653A.114S,Gonz2022A&A...658A.138G}. Current observations show that most planets orbiting around M-dwarfs have masses equivalent to a few Earth to Neptune masses \citep{trifonov2020carmenes,Gonz2022A&A...658A.138G}, while Jupiter-mass planets are observed more frequently around Sun-like or more massive stars \citep{fischer2005planet,reffert2015precise}. GJ 1148, GJ 876 \citep{rivera2010lick,nelson2016empirically,millholland2018new}, and GJ 317 \citep{johnson2007new} are three unusual M-dwarf systems that have massive planets, whose formation is difficult to explain by the core accretion theory of planet formation \citep{Laughlin2004ApJ...612L..73L,morales2019giant}. Alternative formation mechanisms for massive planets around M-dwarfs may be needed \citep{Schlecker2022arXiv220512971S}.

\begin{figure*}
    \centering
    \gridline{\fig{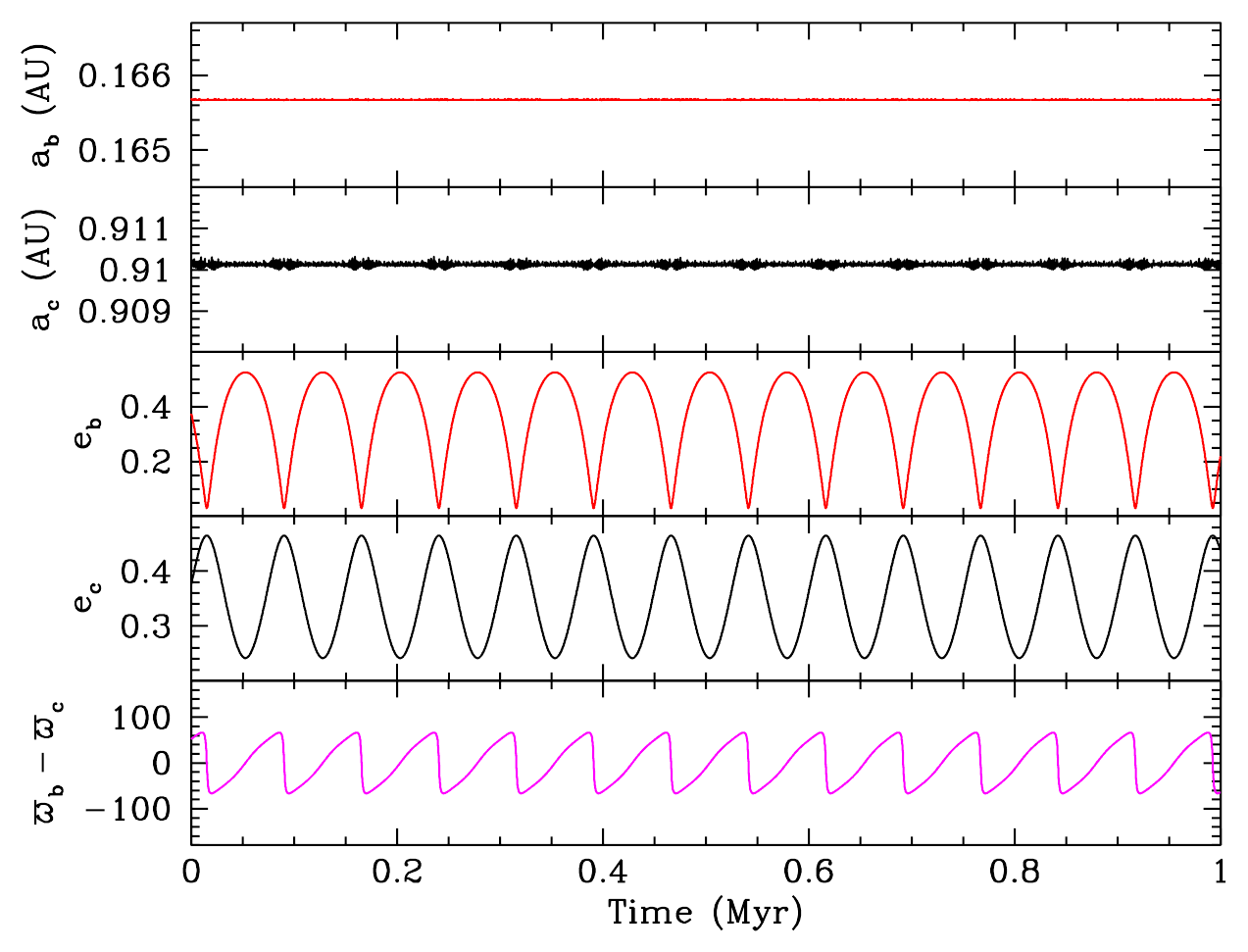}{0.48\textwidth}{}
            \fig{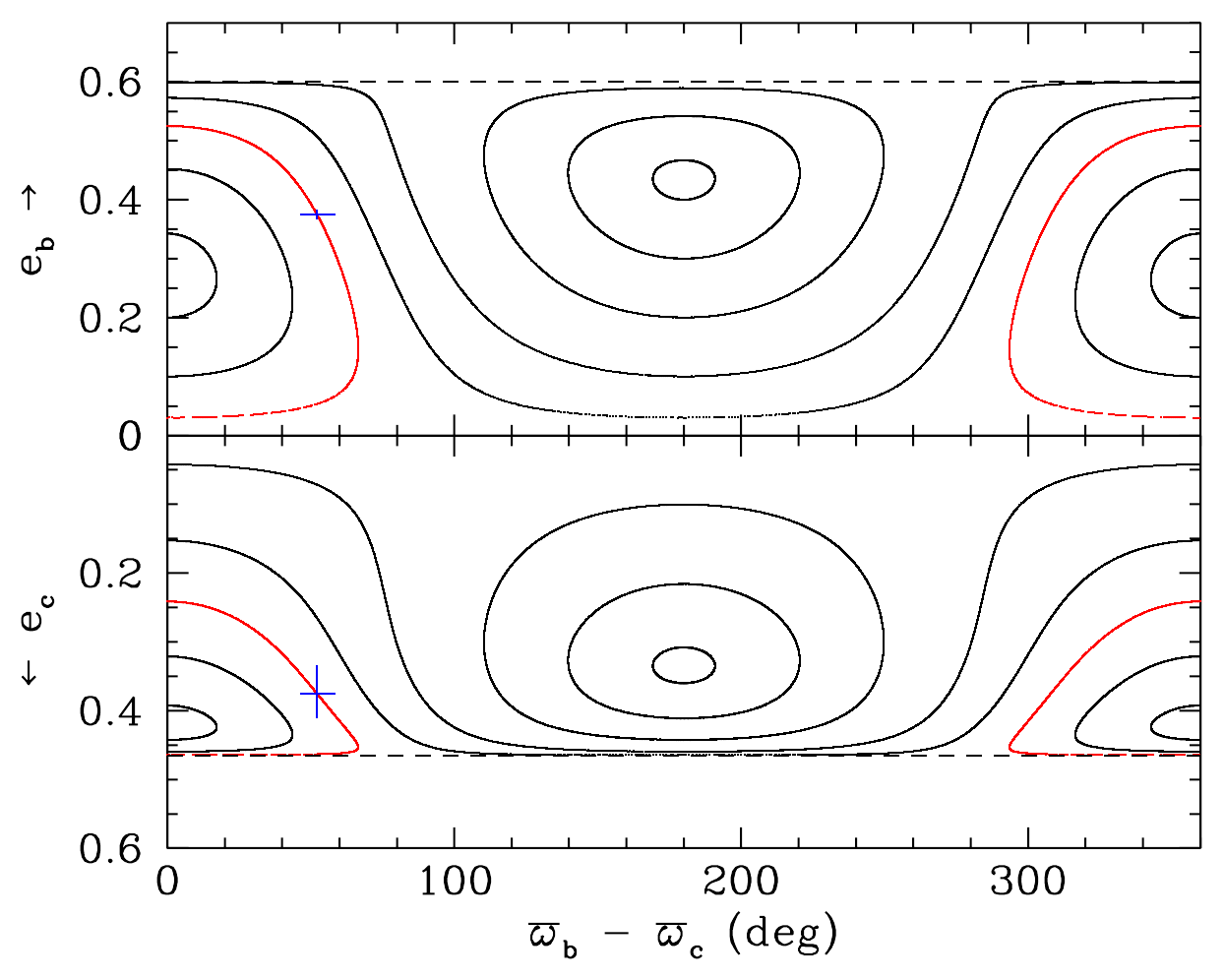}{0.46\textwidth}{}
    }
    \caption{ 
    \textit{Left panel}: Orbital evolution of the best fit of the GJ 1148 system over 1 Myr \citep{trifonov2020carmenes}. \textit{Right panel}: Phase-space diagrams of the planetary eccentricities $e_{b}$ and $e_{c}$ versus the apsidal alignment angle $\Delta \varpi = \varpi_{b} - \varpi_{c}$ for systems with the same total angular momentum as the GJ 1148 system. Blue crosses indicate the best-fit parameters of $e_{b}$, $e_{c}$ and $\varpi_{b} - \varpi_{c}$ and their uncertainties, while the red paths represent the trajectories starting at the best fit.}
    \label{fig:GJ1148_orbital_evolution_Myr}
\end{figure*}

GJ 1148 is an M-type dwarf hosting two Saturn-mass planets \citep{trifonov2018carmenes,trifonov2020carmenes}. The star has mass $M_\ast = 0.354 \pm 0.015$ \(M_\odot\), effective temperature $T_{\text{eff}}$ = $3358 \pm 51$ K, and luminosity $L$ = $0.0143 \pm 0.0003$ \(L_\odot\) \citep{passegger2018carmenes,schweitzer2019carmenes}. The planet GJ 1148 b, with minimum mass $m_{b} \sin{i} = 0.304 M_{\text{J}}$ (where $M_{\text{J}}$ is Jupiter mass), semimajor axis $a_{b}$ = 0.166 au, and orbital period of around 41.4 days, was discovered based on 37 HIRES RV measurements by \cite{haghighipour2010lick}. With an additional 52 RV data from CARMENES, \cite{trifonov2018carmenes} found the second massive planet GJ 1148 c with a minimum mass of $m_{b} \sin{i} = 0.227 M_{\text{J}}$ located in a relatively distant orbit with semimajor axis $a_{c}$ = 0.910 au and orbital period of 532.6 days. The period ratio between the two planets is around 13, and both have orbital eccentricities of around 0.375 at the current epoch. Long-term integration shows that the two planets have large eccentricity variations on a secular timescale of around $72,000\,$yr, as shown in Figure \ref{fig:GJ1148_orbital_evolution_Myr}.

The inner planet GJ 1148 b is in the regime of warm Jupiters (WJs), which are commonly defined as Jupiter-size planets with orbital periods of 10--200 days (e.g., \citealt{dawson2018}).\footnote{
This definition of WJ is primarily for Jupiter-size planets around Sun-like stars. It is unclear in the literature how the definition of WJ should change with stellar mass and luminosity. If we assume that the orbital period range of 10--200 days is for a star of $1\,M_\odot$ and $1\,L_\odot$, the corresponding range in zero-albedo equilibrium temperature is $923$--$340\,$K. The zero-albedo equilibrium temperature of GJ 1148 b is lower at $236\,$K. On the other hand, GJ 1148 b is dynamically similar to eccentric WJ with external giant planet companions.
}
Observations show that WJs can be classified into two populations \citep{sun2022highly} according to their eccentricity and companionship, whose origins are still hotly debated. The WJs with low eccentricity and inner super-Earth companions are suggested to form \textit{in situ} \citep{huang2016warm}. The WJs with moderate eccentricity and Jovian companion are proposed to form through high-eccentricity migration \citep{dawson2014class,dong2013warm,petrovich2016warm}. However, \cite{antonini2016dynamical} found that most observed WJs with characterized external giant planet companions cannot be formed in this way. Another source of eccentricity excitation is planet-planet scattering \citep{rasio1996dynamical,weidenschilling1996gravitational,marzari2002eccentric,chatterjee2008dynamical,juric2008dynamical,raymond2010planet}. \cite{anderson2020situ} studied a scenario of \textit{in situ} scatterings. All of their two-planet systems arise from planet-planet collision events, which produce too many low-eccentricity and closely packed two-planet systems. The dearth of stable two-planet systems from ejection events in \cite{anderson2020situ} is very surprising and even doubtful, because previous studies of three-planet scatterings at several au have shown that an ejection event could produce two planets on stable and well-separated orbits \citep{marzari2002eccentric,chatterjee2008dynamical}, making planet ejection a plausible way to produce systems like GJ 1148.

In this study, we aim to test whether the current GJ 1148 system could have evolved from a tightly packed system of three Saturn-mass planets by losing one of the planets through planet-planet scattering. Previously scattering studies always assumed a Sun-like star with Jupiter-mass planets. Our study explores a new parameter space: Saturn-mass planets around an M-dwarf star, although the planet-star mass ratio is comparable to Jupiter around a Sun-like star. Studying the origin of such a system could provide important implications for the formation of massive planets around M-dwarfs, i.e., it would give the possible parameter space for the configuration of the original GJ 1148 system just after the gas disk dispersal. The success in reproducing the GJ 1148 system with a planet-planet scattering scenario also gives important implications for WJs with a distant companion.

This paper is organized as follows. In Section \ref{sec:Simulations}, we introduce the setups for our $N$-body simulations. In Section \ref{sec:Results}, we show the outcomes of scattering, illustrate the properties of two-planet systems from either planet-planet collision or ejection events, and determine the optimal original configuration of the GJ 1148 system. In Section \ref{sec:Discussion}, we discuss the influence of general relativistic (GR) apsidal precession and the mass of the ejected planet, as well as the discrepancy with the simulations of \textit{in situ} scattering of WJs by \cite{anderson2020situ}. We end with conclusions in Section \ref{sec:Conclusion}.

%%%%%%%%%%
\section{Initial Conditions and Parameters of Simulations} \label{sec:Simulations}

We assume that the original GJ 1148 system includes an M-dwarf star ($M_{\star}$ = 0.354 $M_{\odot}$) and three giant planets. Two planets have the same mass as GJ 1148 b (0.304 $M_{\text{J}}$) and GJ 1148 c (0.227 $M_{\text{J}}$). The third planet is considered a less-massive planet with 0.1 $M_{\text{J}}$. Such a mass choice is consistent with previous scattering studies \citep{chatterjee2008dynamical,anderson2020situ}, which indicate that the planet with the lowest mass is most likely to be ejected. The planets are randomly ordered. Considering all planets are young, the planetary radius is set as 1.0 $R_{J}$. The initial conditions of planet orbits are similar to \cite{anderson2020situ}. For each planet, we sample the argument of periapse, longitude of the ascending node, and mean anomaly in [$0^{\circ}$, $360^{\circ}$], eccentricity in [0.01, 0.05], and inclination in the range of [$0^{\circ}$, $2^{\circ}$], in a uniform random manner. The planets are separated by $K$ in units of mutual Hill radius, which is defined as 
\begin{equation} \label{RHm_eq}
 R_{m,\,H} = \left( \frac{m_{j}+m_{j+1}}{3M_{*}}\right) ^{1/3} \frac{a_{j}+a_{j+1}}{2}.
\end{equation} To study the influence of initial separation between planets, we perform simulations in three groups with $K$ equal to 3.5, 4, and 4.5. In each group, we firstly perform $N_{\text{total}} = 200$ trial $N$-body simulations (\textit{Sim.\ trial}) with the initial innermost planet's semimajor axis $a_{i,1}$ in the range of [0.1, 0.5] au to get a preliminary knowledge of the statistics of scattering outcomes (see Table \ref{t:sims}). Subsequently, two additional sets of simulations (\textit{Sim.\ 1} and \textit{Sim.\ 2}), are performed; they are also listed in Table \ref{t:sims} and will be discussed in Sections \ref{subsec:2p-ej} and \ref{subsec:search_best}. 

\begin{deluxetable*}{lrccl}
\tabletypesize{\scriptsize}
\tablewidth{0pt}
\tablecaption{Parameters of Different Sets of Simulations
\label{t:sims}}
\tablehead{
\colhead{Name} & \colhead{$N_{\text{total}}$} & \colhead{$a_{i,1}$ (au)} & \colhead{$K$} & \colhead{Integration to End of Phase 2}
}
\startdata
{\it Sim.\ trial} & 200  & 0.1 -- 0.5  & 3.5, 4.0, 4.5 & All systems with more than one planet\\
{\it Sim.\ 1}     & 1000 & 0.1 -- 0.5  & 3.5, 4.0, 4.5 & Only two-planet systems from ejection of planet with $0.1 M_{\text{J}}$\\
{\it Sim.\ 2}     & 800  & 0.15 -- 0.3 & 4.5 & Only two-planet systems with $0.304 M_{\text{J}}$ inner planet and $0.227 M_{\text{J}}$ outer planet\\
\enddata
\end{deluxetable*}

We integrate the planetary orbits with the symplectic integrator SyMBA \citep{duncan1998multiple}, which can handle close encounters between the planets. To ensure that SyMBA can adequately integrate the close orbit of the innermost planet,  we set the time step as 3.5 $\times 10^{-4}\,$yr. The initial three-planet system is dynamically unstable. As the system evolves, the growing instability can lead to close encounters between planets. During a close encounter between two planets, the dynamical outcome depends on the relative magnitude between the escape velocity $V_{\text{esc}}$ from the planet’s surface and its orbital velocity $V_{\text{orb}}$, which is quantified by the Safronov number \citep{morbidelli2013dynamical}:
\begin{equation} \label{stab_condition}
 \Theta = \frac{1}{2} \left( \frac{V_{\text{esc}}}{V_{\text{orb}}} \right)^{2} = \left(\frac{M_{p}}{M_{*}}\right) \left(\frac{R_{p}}{a_{p}}\right)^{-1},
\end{equation}where $a_{p}$ is the semimajor axis, $R_{p}$ is the radius of the planet, and $M_{p}$ and $M_{*}$ are the mass of the planet and the host star, respectively. If $\Theta$ is much less than unity, planet-planet collisions will dominate the dynamical outcomes. If $\Theta$ is much greater than unity, planet ejections are favored. Our initial planets have $\Theta$ smaller than unity (with $\Theta = 0.057$--$0.877$), and thus primarily result in planet-planet collisions, followed by planet ejections and planet-star collisions. When the distance between two planets is less than the sum of their radii, they are merged together to form a new planet with the sum of the masses. A planet is ejected and removed when its distance from the host M-dwarf exceeds 1000 au. We also set an inner boundary at 0.02 au to remove planets that come too close to the star. The removed planet would probably result in a collision with the star or survive in an eccentric orbit with an extremely small periastron distance, in which additional physical processes like tidal effect or mass loss would be nonnegligible.

Each three-planet system is firstly integrated with a maximum time $2\pi \times 10^{6}$ $P_{i}$ (except for systems initially separated by $K = 4.5$, for which we set the maximum time as $2\pi \times 10^{7}$ $P_{i}$), with $P_{i}$ the initial orbital period of the initial innermost planet. We refer to this highly chaotic phase as \textit{Phase 1}. The instability often leads to orbital crossings and close encounters and results in a combination of collision, ejection, and removal at the inner boundary. Most systems end up with two planets left. For \textit{Sim.\ trial}, we integrate systems with more than one planet for an additional $2\pi \times 10^{8}$ $P_{in}$, with $P_{in}$ the inner planet’s period at the end of \textit{Phase 1}. This long-term integration is referred to as  \textit{Phase 2}. For \textit{Sim.\ 1} and \textit{Sim.\ 2}, only some of the two-planet systems are integrated to the end of \textit{Phase 2} (see Table \ref{t:sims}).

%-------------------------------------
%%%%%%%%%%%% Results %%%%%%%%%%%%%%%%%%%
%--------------------------------------
\newpage

\section{Results} \label{sec:Results}
\subsection{Scattering outcomes from \textit{Sim.\ trial}}

The instabilities in initially unstable three-planet systems lead to close encounters, which result in planet loss by planet-planet collision (pp), planet ejection (ej), and planet-star close approach (ps). We first made a trial simulation (\textit{Sim.\ trial}) to get a preliminary knowledge of the evolutionary outcomes of the unstable three-planet systems. In each group of $K = 3.5$, 4.0, and 4.5, we performed 200 simulations with time covering both \textit{Phase 1} and \textit{Phase 2}. Almost all of our three-planet systems lose at least one planet at the end of \textit{Phase 2}, (except for five systems in the simulation set with $K = 4.5$, as shown in Table \ref{t:ScaterOutcomesGroup1}) and evolve into one-planet or two-planet systems at the end of \textit{Phase 2}.

\begin{figure}[t]
    \centering
    \includegraphics[width=.5\textwidth]{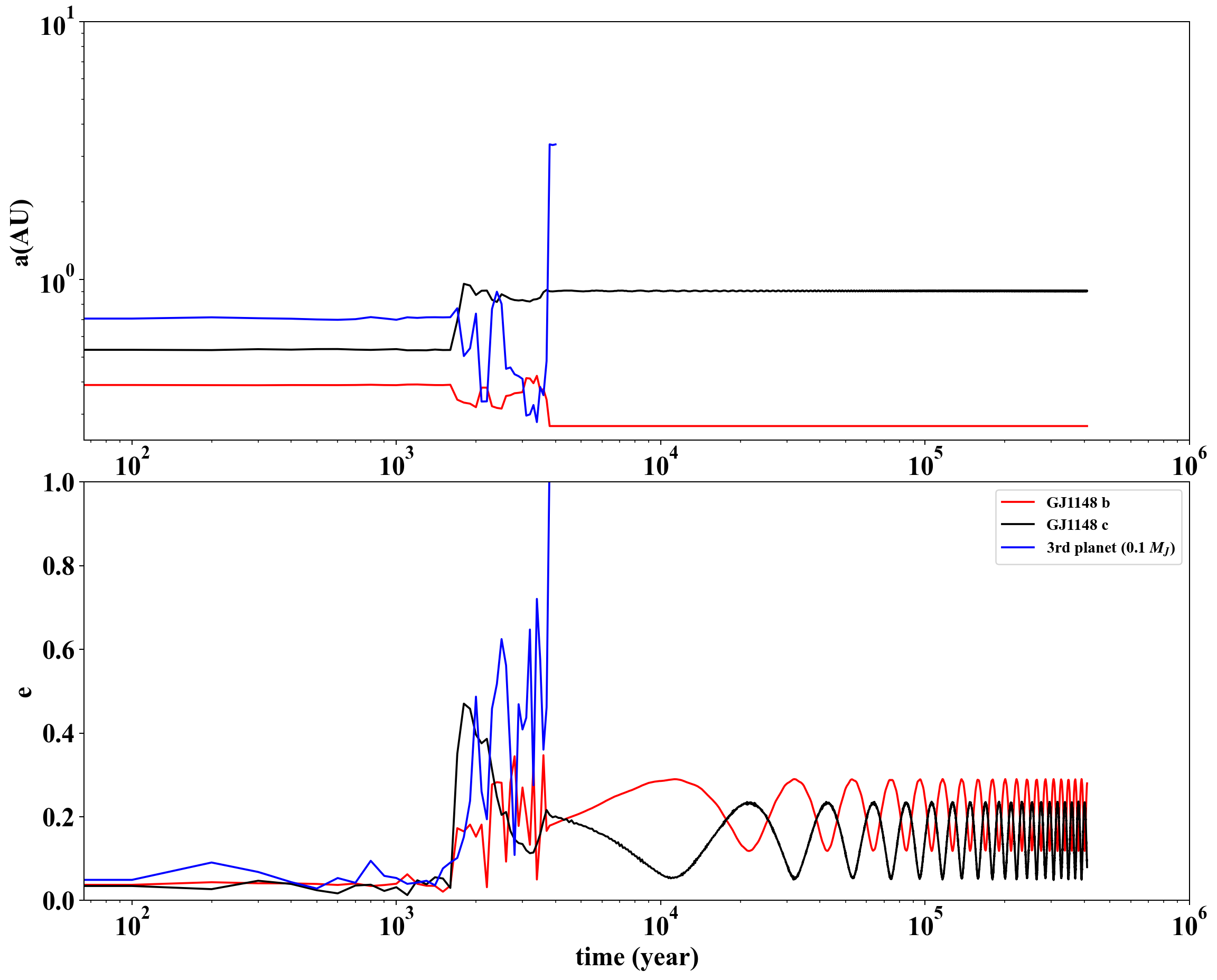}
    \caption{Dynamical evolution of a hypothetical three-planet system with $K = 4.0$ that evolves to a two-planet system. Red, black, and blue colors denote GJ 1148 b, c, and the third planet, respectively. The hypothetical third planet is ejected at around $4 \times 10^{3}\,$yr.}
    \label{fig:chap2:2P_2_4RH_a_e_LongTerm_evolution} 
\end{figure}

Figure \ref{fig:chap2:2P_2_4RH_a_e_LongTerm_evolution} shows the orbital evolution of a system with $K = 4.0$, ending with two planets in stable orbits. The planets in the initial system were GJ 1148 b, c, and the hypothetical third planet of 0.1 $M_{\text{J}}$ in order of semimajor axis. The innermost planet's semimajor axis $a_{i}$ is around 0.4 au. After about $1.3 \times 10^{3}\,$yr, there is a chaotic phase in which orbital crossings repeatedly occur until the third planet is ejected. The original inner planet is scattered closer to the star, while the original middle one is left in a larger orbit. Their orbits are stable, and no orbital crossing occurs. The lower panel shows the time evolution of the eccentricities. We can see that the eccentricities change with small amplitudes before $1.3 \times 10^{3}\,$yr. During the chaotic phase, the eccentricities of all planets are significantly excited. The eccentricity of the third planet increases to exceed unity and the planet ends up being ejected. In the final stable phase, we can see that the eccentricities of the two planets vary on a secular time scale due to their gravitational interactions. The final configuration with two planets that are widely separated and show moderately large eccentricity variations is qualitatively similar to the GJ 1148 system.

\begin{figure}[t]
    \centering
    \includegraphics[width=.5\textwidth]{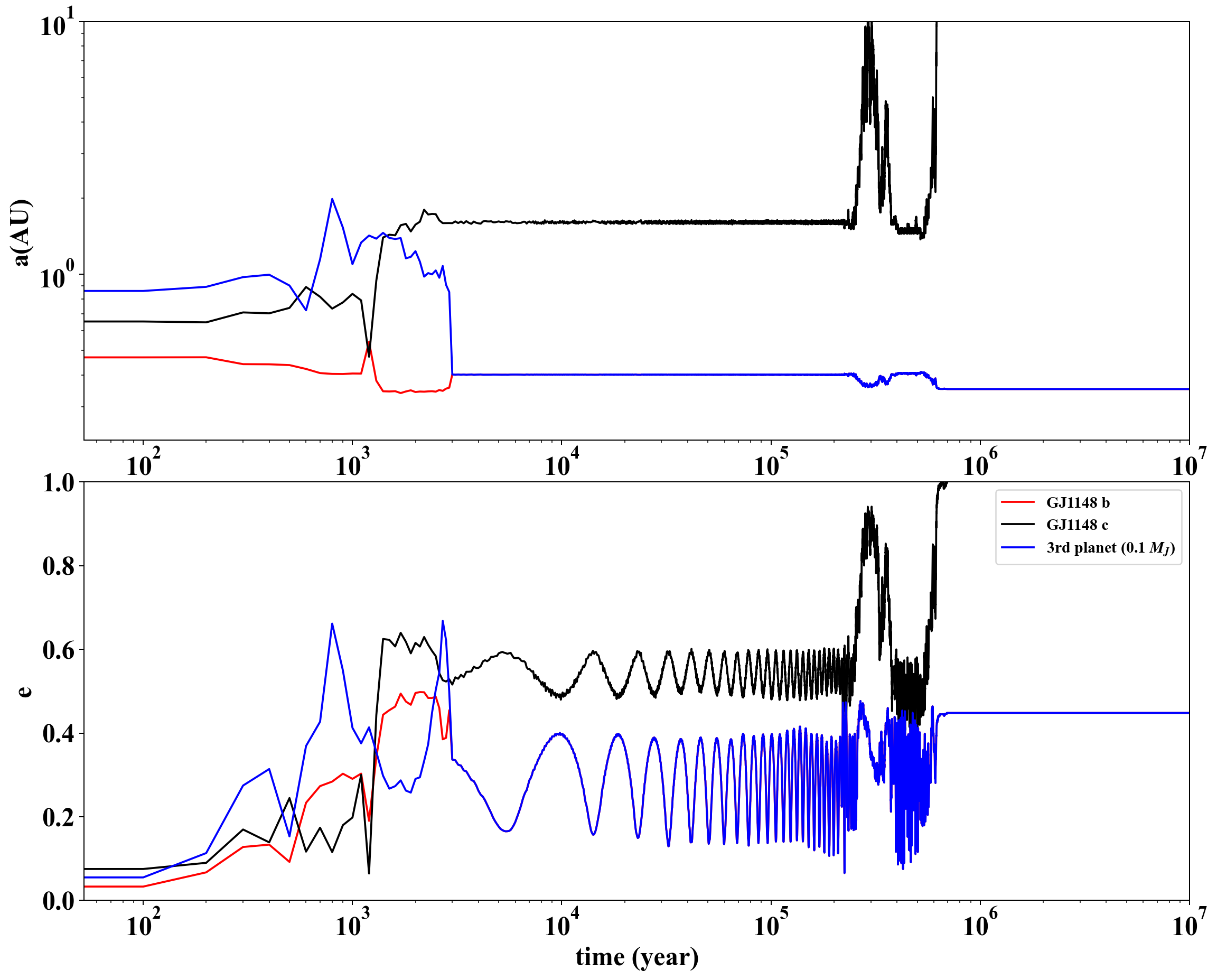}
    \caption{Similar to Figure \ref{fig:chap2:2P_2_4RH_a_e_LongTerm_evolution}, but for a hypothetical three-planet system with $K = 4.0$ that ends up with only one planet left. Note that after the merger of GJ 1148 b and the third planet at $\simeq 3 \times 10^{3}\,$yr, the blue curve denotes the new and more massive planet.}
    \label{fig:chap2:1P_13_4RH_a_e_LongTerm_evolution}
\end{figure}

Figure \ref{fig:chap2:1P_13_4RH_a_e_LongTerm_evolution} shows the time evolution of a system that loses two planets. The three planets start to evolve chaotically after a period without orbital crossing. During the chaotic phase, the planetary eccentricities and semimajor axes change significantly. At a time of $\simeq 3 \times 10^{3}\,$yr, the original outermost planet collides with the original innermost planet, merging into a new planet in a slightly smaller orbit than the initial innermost orbit. The original middle planet is scattered to a larger eccentric orbit at the end of the chaotic phase. The orbits then evolve slowly until $\simeq 2 \times 10^{5}\,$yr, when the semimajor axis and eccentricity of the outer planet suddenly increase, while the orbit of the inner planet changes more mildly. The outer planet is finally ejected at a time $\simeq 7  \times 10^{5}\,$yr, making the inner planet’s eccentricity larger and its semimajor axis smaller.

\begin{figure*}[t!]
    \centering
    \includegraphics[width=1\textwidth]{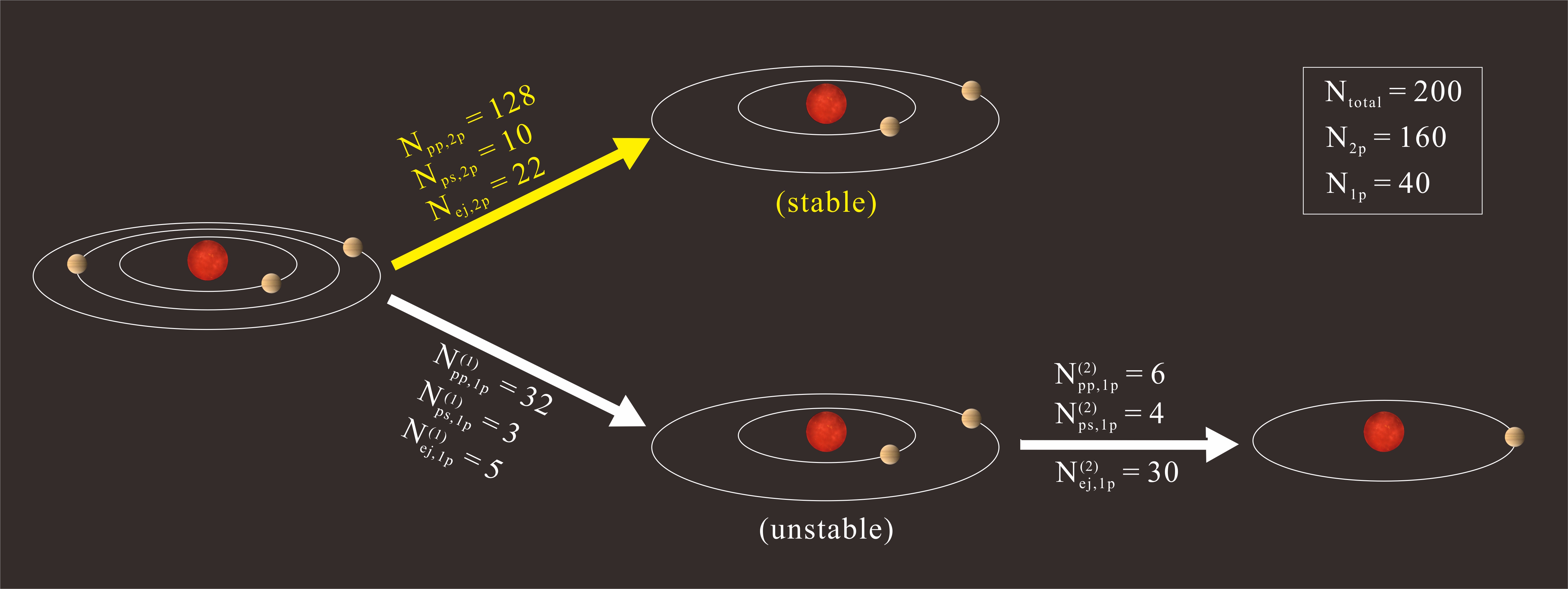}
    \caption{Statistics of initially unstable three-planet systems evolving into two-planet systems and one-planet systems in trial simulations (\textit{Sim.\ trial}) with $K = 4.0$ (see also Table \ref{t:ScaterOutcomesGroup1}).} 
    \label{fig:chap2:brachratio4RH}
\end{figure*}

Figure \ref{fig:chap2:brachratio4RH} shows the branching ratio of how initially unstable three-planet systems evolve into two-planet and one-planet systems for $K$ = 4.0 simulations. The most common events that result in the first planet loss are planet-planet collisions ($N_{pp,2p} + N_{pp,1p}^{(1)} = 160$), followed by planet ejections ($N_{ej,2p} + N_{ej,1p}^{(1)} = 27$) and planet-star close approach ($N_{ps,2p} + N_{ps,1p}^{(1)} = 13$). The scattering outcomes dominated by collisions are consistent with the small value of the Safronov number $\Theta$ discussed in Section \ref{sec:Simulations}. For the second-planet loss, the most common events are planet ejections ($N_{ej,1p}^{(2)} = 30$), followed by planet-planet collisions ($N_{pp,1p}^{(2)} = 6$) and planet-star collisions ($N_{ps,1p}^{(2)} = 4$). At the end of \textit{Phase 2}, the total number of final stable one- and two-planet systems are $N_{F,\,1P} = N_{pp,1p}^{(2)} + N_{ps,1p}^{(2)} + N_{ej,1p}^{(2)} = 40$ and $N_{F,\,2P} = N_{pp,2p} + N_{ps,2p} + N_{ej,2p} = 160$, respectively.

\begin{table*}[tb]
\scriptsize
\tablewidth{0pt}
\centering
\caption{Scattering Outcomes of Trial Simulations (\textit{Sim.\ trial}) Starting with Different Initial Separations}
\label{t:ScaterOutcomesGroup1}
\begin{tabular}{lccccccccccccc}
\hline \hline
$K$ & $N_{\text{total}}$ & $N_{pp,2p}$ & $N_{ps,2p}$& $N_{ej,2p}$ & $N_{pp,1p}^{(1)}$ & $N_{ps,1p}^{(1)}$ & $N_{ej,1p}^{(1)}$ & $N_{pp,1p}^{(2)}$ & $N_{ps,1p}^{(2)}$ & $N_{ej,1p}^{(2)}$ & $N_{F,\,1P}$ & $N_{F,\, 2P}$ & $N_{F,\,3P}$\\

\hline

3.5 & 200 & 126 & 1  & 17 & 47 & 3 & 6 & 1 & 8 & 47 & 56 & 144 & 0 \\
4.0 & 200 & 128 & 10 & 22 & 32 & 3 & 5 & 6 & 4 & 30 & 40 & 160 & 0 \\
4.5 & 200 & 145 & 7  & 22 & 12 & 1 & 8 & 2 & 1 & 18 & 21 & 174 & 5 \\
\hline
\end{tabular}
\tablecomments{The first two columns are the initial separation between the planets in units of mutual Hill radii ($K$) and the number of simulations ($N_{\text{total}}$). The quantities $N_{pp,2p}$, $N_{ps,2p}$, $N_{ej,2p}$ are the numbers of two-planet (2p) systems from planet-planet collision (pp), planet-star close approach (ps), and planet ejection (ej), respectively. The quantities $N_{pp,1p}^{(1)}$, $N_{ps,1p}^{(1)}$, and $N_{ej,1p}^{(1)}$, $N_{pp,1p}^{(2)}$, $N_{ps,1p}^{(2)}$, and $N_{ej,1p}^{(2)}$ are similar, but for the loss of the first (superscript 1) and second (superscript 2) planets of one-planet (1p) systems.
The last three columns are the numbers of final one-planet ($N_{F,\, 1P}$), two-planet ($N_{F,\, 2P}$), and three-planet ($N_{F,\, 3P}$) systems.}
\end{table*}

The outcomes of simulations (\textit{Sim.\ trial}) with different $K$ are summarized in Table \ref{t:ScaterOutcomesGroup1}. As expected, the $K = 3.5$ simulations produce the highest number of planet-planet collisions since the three planets' mean $\Theta$ value is the smallest. We also notice that both $K = 3.5$ and 4 simulations do not have three-planet systems left at the end of integration, but five three-plant systems are left in $K = 4.5$ simulations. This is also not surprising because dynamical instabilities take a long time to grow for a large initial separation of $K = 4.5$. In addition, we observe that the fraction of final one-planet systems seems to decrease with $K$. It indicates that systems with larger $K$ are less likely to experience second-planet loss. 

The two-planet systems that follow a planet-star close approach (ps) are not studied here because the planet is removed artificially, and the properties of the remaining two planets are not realistic. We focus on the remaining two-planet systems following an ejection (post-ejection two-planet system) or a collision event (post-collision two-planet system). We will show their properties and compare them with the observation of the GJ 1148 system in the following.  

%%%%%%%%%%%%%%% 2P_PP %%%%%%%%%%%%%%%%%%%%%%%%%
\subsection{Properties of post-collision two-planet systems from \textit{Sim.\ trial}} \label{subsec:2p-pp}

\begin{figure*}[ht!]
\centering
\includegraphics[width=1\textwidth]{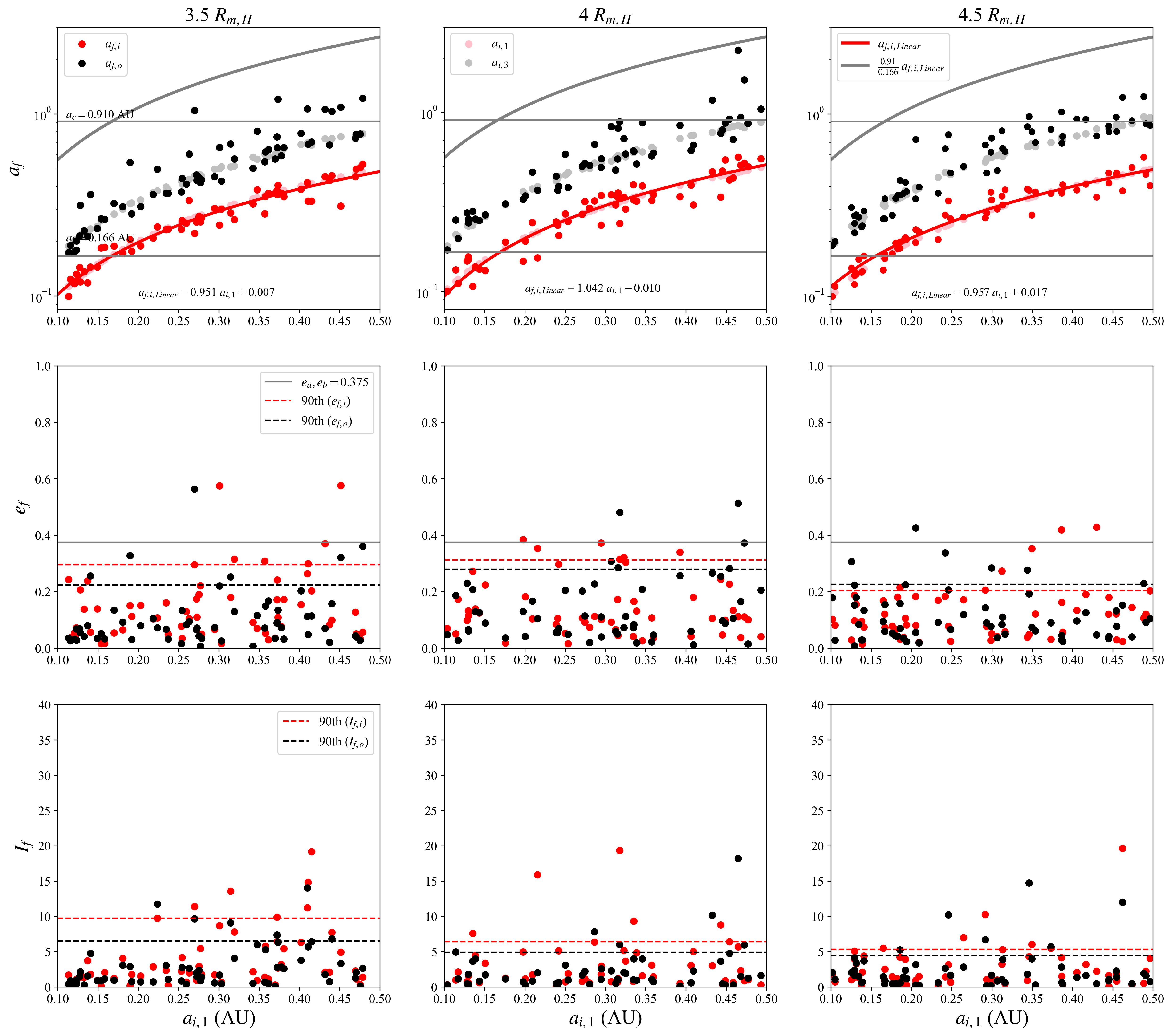}
 \caption{Properties of post-collision two-planet systems for simulations with initial separation $K$ = 3.5, 4, and 4.5. Only systems from the first 100 simulations are shown for clarity.}
    \label{fig:chap2:2P_PP_difK}
\end{figure*} 

Figure \ref{fig:chap2:2P_PP_difK} shows another view of the properties of the post-collision two-planet systems for simulations with initial spacing $K = 3.5$, 4.0, and 4.5. The upper panels show the semimajor axes of the remaining inner planets $a_{f,\, i}$ (red) and outer planets $a_{f,\, o}$ (black) versus $a_{i,\,1}$. The semimajor axes of the initial innermost $a_{i,\,1}$ (pink) and outermost $a_{i,\,3}$ (grey) planets are included for comparison. We can see that a linear relationship can fit the semimajor axis of the remaining inner planet ($a_{f,\, i}$) versus the semimajor axis of the initial innermost planet ($a_{i,\,1}$) for all three sets of simulations very well. The semimajor axis $a_{f,\, i}$ (red circles) shows a small scatter around $a_{i, 1}$ (pink circles), and the slopes of the three linear models from fitting $a_{f,\, i}$ versus $a_{i, 1}$ are close to unity, showing that the position of the inner planet almost remains the same. The semimajor axis $a_{f,\,o}$ seems to have a wider range of variations, but it has a tendency to increase with $a_{i,1}$ and is generally close to the semimajor axis of the initial outermost planet $a_{i,3}$. So the positions of the remaining planets do not change much compared to the innermost and outermost planets of the initial three-planet systems. Compared to the observation, we find that the post-collision two-planet systems with the inner planet's semimajor axis around 0.166 au could reproduce the semimajor axis of GJ 1148 b. However, the orbit of the produced outer planet is too small to match the semimajor axis of GJ 1148 c.

The middle panels of Figure \ref{fig:chap2:2P_PP_difK} show the eccentricities ($e_{f,\, i}$ and $e_{f,\, o}$) of the remaining two planets versus $a_{i,1}$. The horizontal grey line shows the observed eccentricity of 0.375 for both planets \citep{trifonov2020carmenes}. The red and black dashed lines represent the 90th percentiles of $e_{f,\, i}$ and $e_{f,\, o}$, respectively, which are lower than the grey line, showing that post-collision two-planet systems have lower eccentricities compared to the observation. The property of low eccentricities for the planets after a collision is consistent with previous studies \citep{ford2001dynamical,anderson2020situ}. Nevertheless, the eccentricities of GJ 1148 b and c are not constant but vary with large amplitude due to their secular interaction with each other. So, low eccentricity is allowed, but only for one of the planets. However, only few post-collision two-planet systems have either $e_{f,\,i }$ or $e_{f,\, o}$ exceeding 0.375. In addition, we observe that, unlike the semimajor axis, there is no evident dependence between the final eccentricities and initial $a_{i,1}$.

The lower panels of Figure \ref{fig:chap2:2P_PP_difK} show the inner and outer planets' inclinations. For $K = 3.5$ simulations, 90\% of inner (outer) planets have orbital inclination below $10^{\circ}$ ($6^{\circ}$). The orbital inclination of the inner (outer) planets in $K = 4.0$ simulations has 90th percentiles around $6^{\circ}$ ($5^{\circ}$). In $K = 4.5$ simulations, the orbital inclinations of both inner and outer planets have 90th percentiles near $5^\circ$, which are slightly lower than those of $K = 4.0$ simulations. We can see that post-collision two-planet systems generally have low orbital inclinations. 

In summary, Figure \ref{fig:chap2:2P_PP_difK} shows that post-collision two-planet systems have smaller orbital spacing and eccentricities, so they cannot produce systems like GJ 1148.

%%%%%%%%%%%%%%% 2P_EJ %%%%%%%%%%%%%%%%%%%%%%%%%
\subsection{Properties of post-ejection two-planet systems from \textit{Sim.\ trial} and \textit{Sim.\ 1}} \label{subsec:2p-ej}

The fraction of post-ejection two-planet systems in \textit{Sim.\ trial} only constitutes a small percentage of the remaining two-planet systems. For example, in simulations with $K = 4.0$, it accounts for approximately $11\%$ (refer to Table \ref{t:ScaterOutcomesGroup1}). Therefore, the fraction with the same planetary mass order as the GJ 1148 system is even smaller. To get more statistics about post-ejection two-planet systems, we performed an additional 1000 simulations (\textit{Sim.\ 1}) and only fully integrated the two-planet systems arising from the ejection of the hypothetical third planet (Table \ref{t:sims}). 

\begin{figure*}[t!]
\centering
\includegraphics[width=1\textwidth]{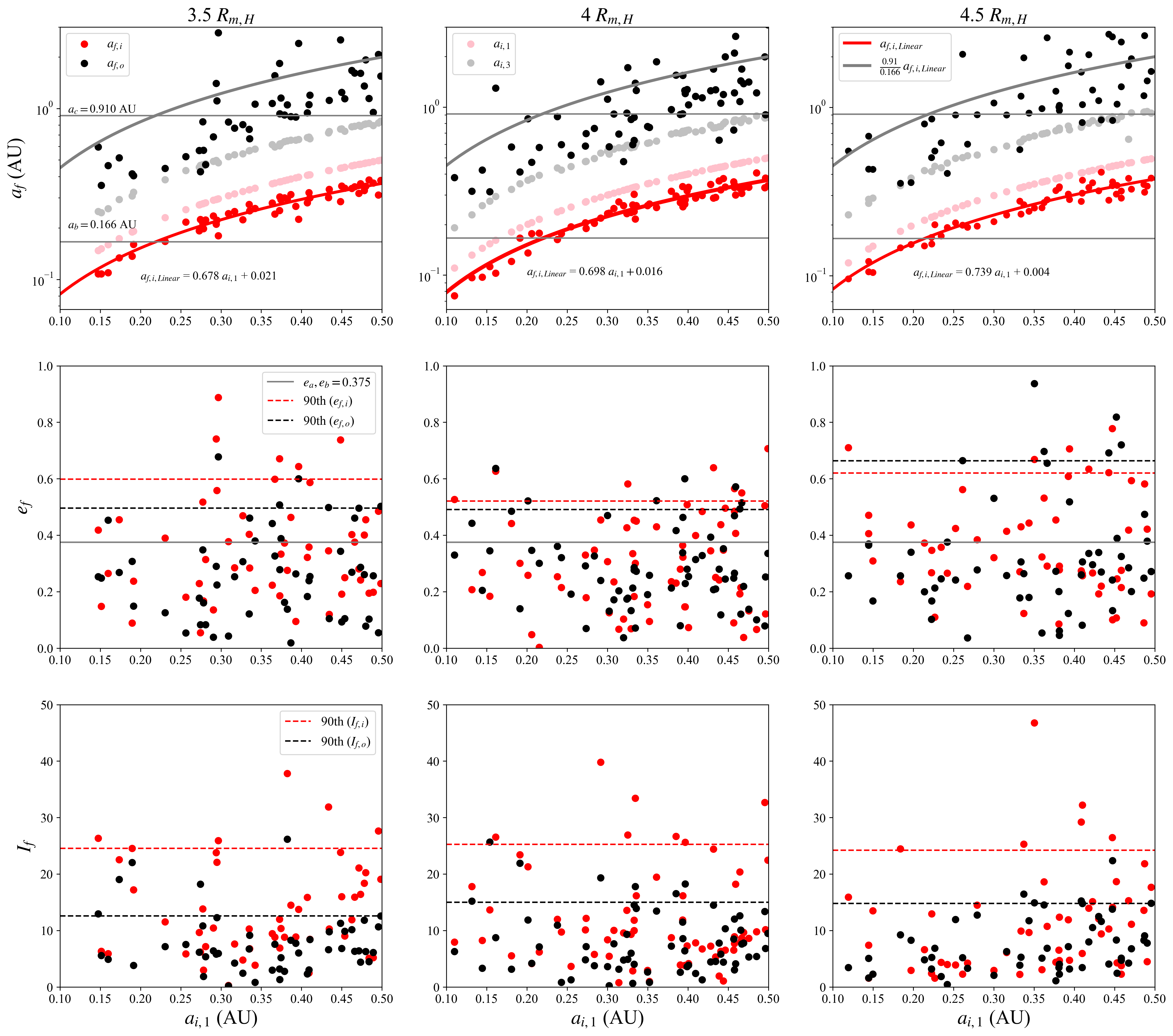}
\caption{Properties of post-ejection two-planet systems (with the same planet mass order as the GJ 1148 system) for simulations (\textit{Sim.\ trial} and \textit{Sim.\ 1}) with initial separation $K$ = 3.5, 4, and 4.5.}
\label{fig:chap2:2P_EJ_difK}
\end{figure*}

Figure \ref{fig:chap2:2P_EJ_difK} shows the orbital properties of the planets in the post-ejection two-planet systems with the same mass order as the GJ 1148 system. The upper panels show that, in each group, $a_{f,\, i}$ versus $a_{i,1}$ can be well fit by a linear relationship, similar to the post-collision cases, but the slope of the linear model is smaller than the post-collision cases. The red circles are lower than the pink circles, showing that $a_{f,\, i}$ is smaller than $a_{i,1}$ and that ejections push the remaining inner planet to a smaller orbit than the initial innermost planet. In addition, we can see the slope of the linear model slightly increases with $K$, indicating that the broader initial systems produce inner planets in orbit closer to the initial inner planets. The semimajor axis $a_{f,\,o}$ also seems to have a linear relationship with $a_{i, 1}$ but with large scatter. The semimajor axis $a_{f,\,o}$ (black circles) is generally above $a_{i,3}$ (grey circles), suggesting that ejection results in the outer planet being in a larger orbit than the initial outermost planet. So, in general, ejection would cause one planet to orbit closer to the star and the other to move to a larger orbit. Comparing simulations with different $K$, we see no major differences in orbital properties. However, at $a_{i,1} \approx$ 0.21 au, which is where $a_{f,i}$ can match the semimajor axis of GJ 1148 b, only $K = 4.0$ and 4.5 simulations have several cases in which $a_{f,o}$ can simultaneously match the semimajor axis of GJ 1148 c.

The middle panels show the eccentricities of the remaining two planets. For all cases, we can see that the distributions of eccentricities do not have an evident correlation with $a_{i,1}$, similar to the post-collision cases. The eccentricities are generally higher than the post-collision cases (with 90th percentiles of eccentricities for both planets exceeding 0.375), which indicates that ejections have significantly excited the eccentricities of the planets. In addition, the 90th percentile of the eccentricity of the inner planet is larger than that of the outer planet in the simulations of $K = 3.5$, while they are nearly equal for $K = 4$ and 4.5. The bottom panels show the inclinations of the two planets following ejections. The inclinations for both planets are higher than those of the post-collision cases in Figure \ref{fig:chap2:2P_PP_difK}, suggesting that ejection also excites inclination. The inclination of the inner planet is higher than that of the outer planet for all $K$. 

Planet-planet collisions produce two-planet systems with small separation and low eccentricities, thus unlikely to reproduce the GJ 1148 system. In contrast, planet ejection tends to produce eccentric and widely separated two-planet systems. The semimajor axis $a_{f,i}$ correlates with $a_{i,1}$, and it can reproduce $a_{b}$ when $a_{i,1}$ is around 0.21 au. Nevertheless, matching $a_{c}$ is not easy because $a_{f,o}$ varies over a wide range. We expect systems like GJ 1148 could be discovered by searching through more post-ejection two-planet systems with $a_{i,1}$ close to 0.21 au.

%%%%%% Searching for the best fit %%%%%%%%%%%%%%%%
\newpage

\subsection{Searching for GJ 1148-like two-planet systems from \textit{Sim.\ 1} and \textit{Sim.\ 2}} \label{subsec:search_best}

\begin{figure*}[!t]
    \centering
    \includegraphics[width=1\textwidth]{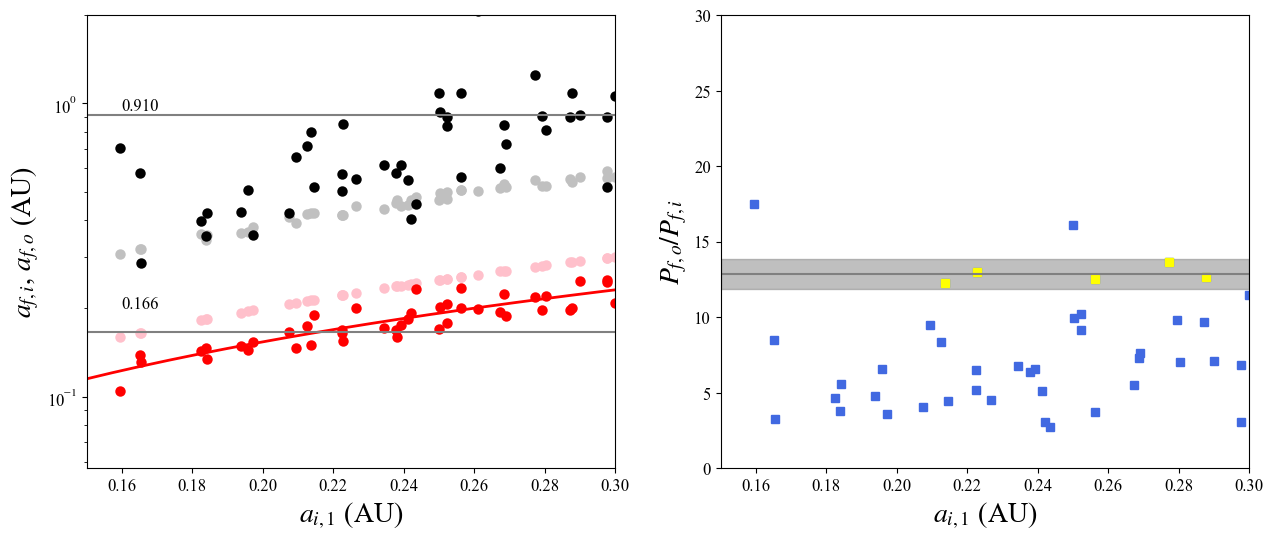}
    \caption{Properties of post-ejection two-planet systems for simulations with initial $a_{i,1}$ in range of [0.15, 0.3] au and separation $K = 4.5$. \textit{Left panel:} $a_{f,\,i}$ and $a_{f,\,o}$ vs.\ $a_{i,1}$ (similar to the upper panels of Figures \ref{fig:chap2:2P_PP_difK} and \ref{fig:chap2:2P_EJ_difK}). \textit{Right panel:} the period ratio $P_{f,\,o}$/$P_{f,\,i}$ vs.\ $a_{i,1}$. The shaded area represents the range in which the difference between the ratio of planetary periods to that of GJ 1148 does not exceed 1. The systems in this area are selected and marked yellow.}    \label{fig:Select_Perioratio13_4d5RH_densesampling}
\end{figure*}

To search for initial configurations of three-planet systems that can produce stable two-planet systems comparable to the GJ 1148 system, we perform \textit{Sim.\ 2}, which includes 800 simulations with $K$ = 4.5 and $a_{i,1}$ in a small range of [0.15, 0.3] au and keeps other setups the same. We only fully integrate the systems in which one planet is ejected, the remaining planet with the mass of GJ 1148 b is in the inner orbit, and the planet with the mass of GJ 1148 c is in the outer orbit. We select systems (including those with $a_{i, 1}$ in the range of [0.15, 0.3] au from \textit{Sim.\ 1}) close to the period ratio (or spacing) of the GJ 1148 system. Then we analyze each system's secular eccentricity variation and apsidal alignment angle $\Delta\varpi$ to select the optimal systems. 

Figure \ref{fig:Select_Perioratio13_4d5RH_densesampling} shows the properties of the post-ejection two-planet systems with $a_{i,1}$ in the range of [0.15, 0.3] au. The left panel shows $a_{f,\,o}$ and $a_{f,\, i}$ compared to the observation, in which we can see several systems with semimajor axes close to both GJ 1148 planets. The right panel shows the period ratio $P_{f,\,o}/P_{f,\,i}$ versus $a_{i,1}$. We find five cases (yellow squares) whose period ratio is within $\pm$1 of GJ 1148's period ratio. Three of them have $a_{i,1}\,>$  0.25 au, which produces two-planet systems with an overall distance farther from the central star than the GJ 1148 planets. We observe that about 24\% of the systems shown in Figure \ref{fig:Select_Perioratio13_4d5RH_densesampling} can reach period ratio $\ga 10$. If the initial system has a more significant separation, such as $5 R_{m,\, H}$, more widely spaced systems close to the GJ 1148 system probably could arise, but the instability would take longer to grow, making the numerical simulations impractical. 

\begin{figure}[!t]
    \centering
    \includegraphics[width=0.45\textwidth]{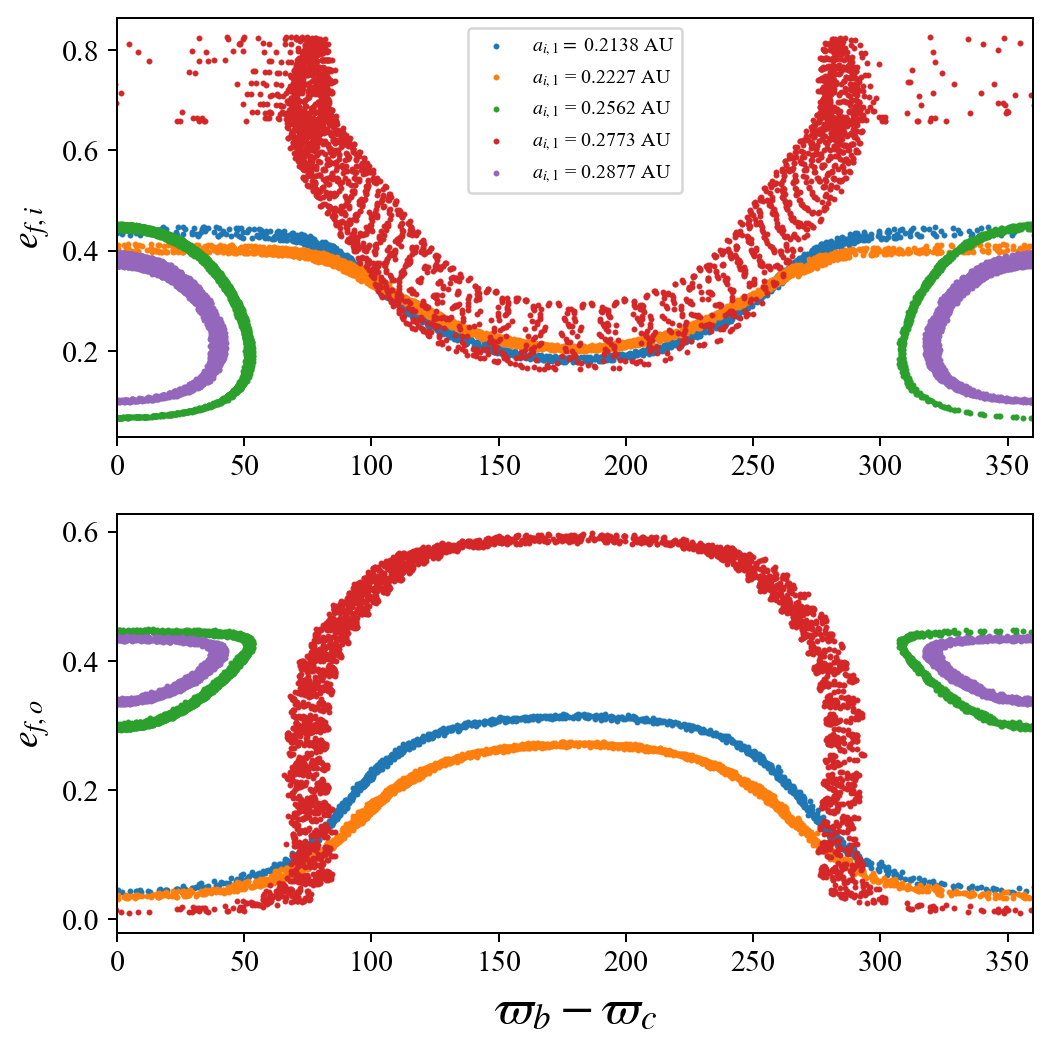}
    \caption{Trajectories of eccentricities versus apsidal alignment angle $\Delta \varpi$ = $\varpi_{b} - \varpi_{c}$ for five selected two-planet systems with comparable period ratios to the GJ 1148 system. The trajectory in green color is the one most similar to the best-fit GJ 1148 system.}   
    \label{fig:pomega_vs_ecc}
\end{figure}

We analyze the five selected planetary systems by comparing their eccentricity  and apsidal alignment angle variations to the GJ 1148 system in Figure \ref{fig:pomega_vs_ecc}. All of the systems are qualitatively similar to GJ 1148 (right panel of Figure \ref{fig:GJ1148_orbital_evolution_Myr}) in showing significant eccentricity variations and $\Delta\varpi$ librating around $0^\circ$ or circulating near the boundary between libration and circulation. We observe that there are three systems with $\Delta \varpi$ circulating and two with $\Delta \varpi$ librating around 0$^{\circ}$. The green one has semi-amplitude of $\Delta \varpi$ around 60$^{\circ}$ and eccentricity variations close to the best-fit GJ 1148 system. We regard this system as the most similar one to the observation found in our scattering experiments.

\begin{figure}[!t]
    \centering
    \includegraphics[width=0.45\textwidth]{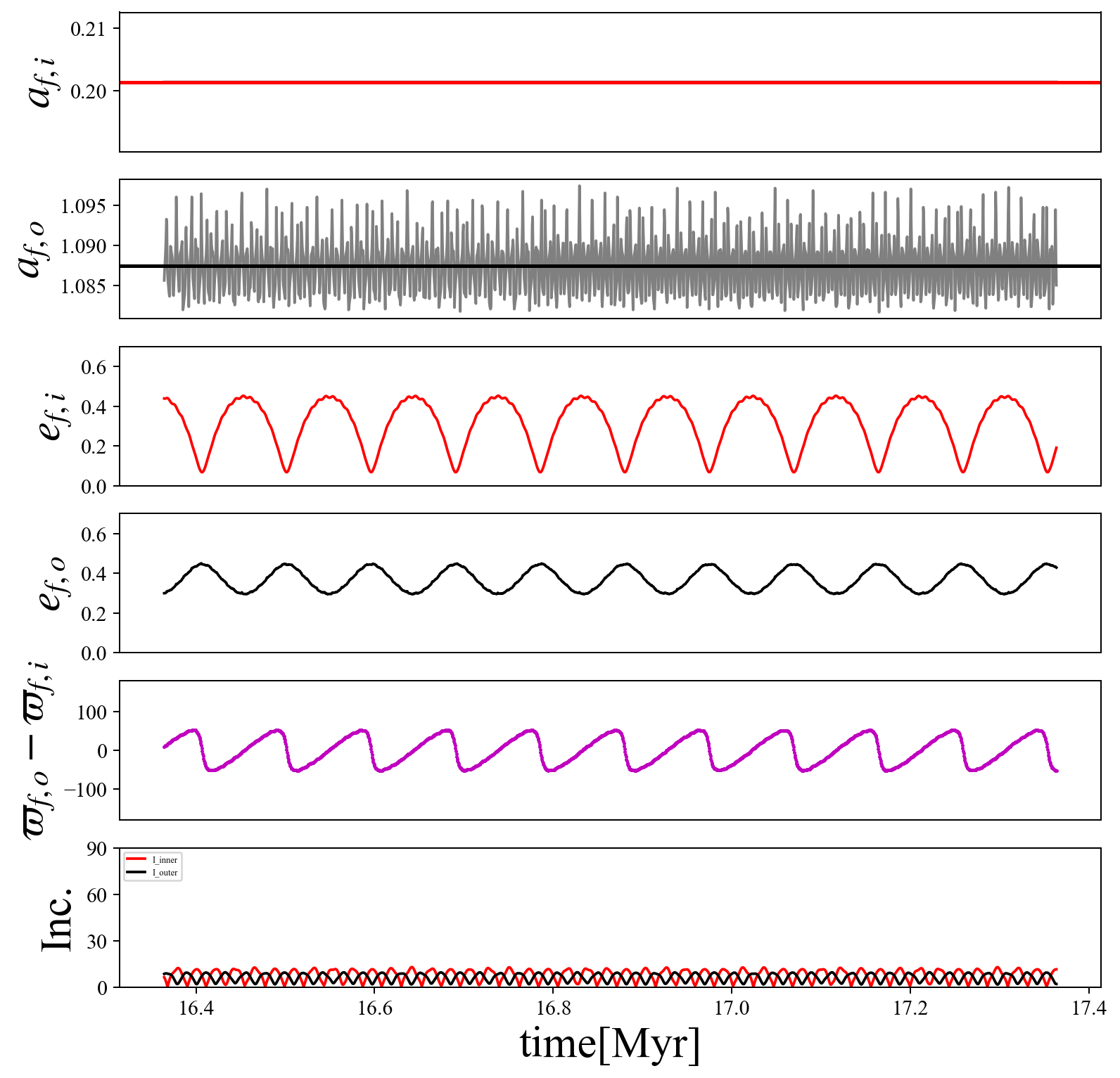}
    \caption{Million-year orbital evolution of a post-ejection two-planet system, which is very similar to the best-fit configuration of the GJ 1148 system (see left panel of Figure \ref{fig:GJ1148_orbital_evolution_Myr}). The time-averaged semimajor axis of its inner planet is 0.201 au (red solid line), and that of the outer planet is about 1.085 au (black solid line). The eccentricity variations of the two planets are very similar to the best-fit model for the GJ 1148 system but on a longer time scale. The apsidal alignment angle $\Delta \varpi$ = $\varpi_{b} - \varpi_{c}$ varies from $-60^{\circ}$ to $60^{\circ}$.}    
    \label{fig:fig_541}
\end{figure}

Figure \ref{fig:fig_541} shows the evolution of this system over a million years, which is similar to the evolution of the best-fit GJ 1148 system shown in the left panel of Figure \ref{fig:GJ1148_orbital_evolution_Myr}. The semimajor axes of the two planets in this hypothetical system do not have significant variation and are generally close to the GJ 1148 system. The semimajor axis of the inner planet is about 0.201 au, which is 0.035 au larger than GJ 1148 b. The mean semimajor axis of the outer planet is about 1.085 au, which is 0.175 au larger than that of the GJ 1148 c. Therefore, the orbital spacing between the two planets in this system is slightly larger than that of the GJ 1148 system. In post-ejection two-planet systems, planets typically exhibit larger inclinations. For instance, the 90th percentile of the inner planet's inclination is approximately $24^\circ$, while that of the outer planet's inclination is roughly $15^\circ$. Nevertheless, the planets in this system have inclinations smaller than $13^\circ$. The eccentricities of the two planets vary with large amplitudes, with $e_{f,\, i}$ from 0.06 to 0.45 and $e_{f, \, o}$ from 0.3 to 0.44, similar to the GJ 1148 planets. We can see that the eccentricities of the two planets are anticorrelated. When $e_{f,\,i}$ reaches its maximum value, $e_{f,\,o}, $ reaches its minimum value. The timescale of the eccentricity variations is about $95,000\,$yr, which is about $23,000\,$yr larger than GJ 1148. The longer timescale is due to the larger orbit semimajor axes of this system. As suggested by Equation (40) in \cite{lee2003secular}, the timescale for secular evolution of eccentricity is proportional to 
\begin{equation} \label{e_secu_timscale}
 t_{e} = \frac{4}{3\alpha^3} \left( \frac{M_{\star}+m_{1}}{m_{2}} \right) \frac{1}{n_{1}},
\end{equation}where $\alpha = a_{1}/a_{2}$ and $n_{1}$ is the mean motion of the inner planet. The $\alpha$ of this system is close to the GJ 1148 system, but $n_{1}$ is smaller than that of GJ 1148 b, making the secular timescale $t_{e}$ longer.  

%%%%%%%%%%%%%%%% Discussion %%%%%%%%%%%%%%%%%%%%%%%%%%%%%
\section{Discussion} \label{sec:Discussion}
\subsection{The influence of GR apsidal precession}

GR apsidal precession has been incorporated in $N$-body codes such as REBOUND \citep{Tamayo2020MNRAS.491.2885T}, which has been used to study the origin of warm Jupiters \citep{anderson2020situ}. However, the effects of GR apsidal precession on planet-planet scattering outcomes have been rarely explored. Here we want to explore whether incorporating GR apsidal precession can produce post-ejection two-planet systems with different properties than without it. We have modified the SyMBA code to include GR apsidal precession by using the potential of \cite{nobili1986simulation}.\footnote{
The source code is available at \url{https://github.com/manhoilee/swift_symba5_GR}
}
The kick to the linear momentum of each planet due to this potential is implemented in barycentric coordinates to conserve total angular momentum. We use the same initial conditions as the $K$ = 4.5 simulations (denoted as $K$4.5) but with GR apsidal precession. We perform a total of 800 simulations (denoted as $K$4.5-GR), the same number as the $K4.5$ simulations in \textit{Sim.\ 2}. As in Section \ref{subsec:search_best}, we only analyze those two-planet systems produced by planet ejection, with GJ 1148 b in an inner orbit and GJ 1148 c in an outer orbit. At the end of Phase 2, only 25 systems remain stable, which is consistent with 31 out of 800 in \textit{Sim.\ 2} of $K4.5$ within statistical uncertainty. 

\begin{figure*}
    \centering
  \gridline{\fig{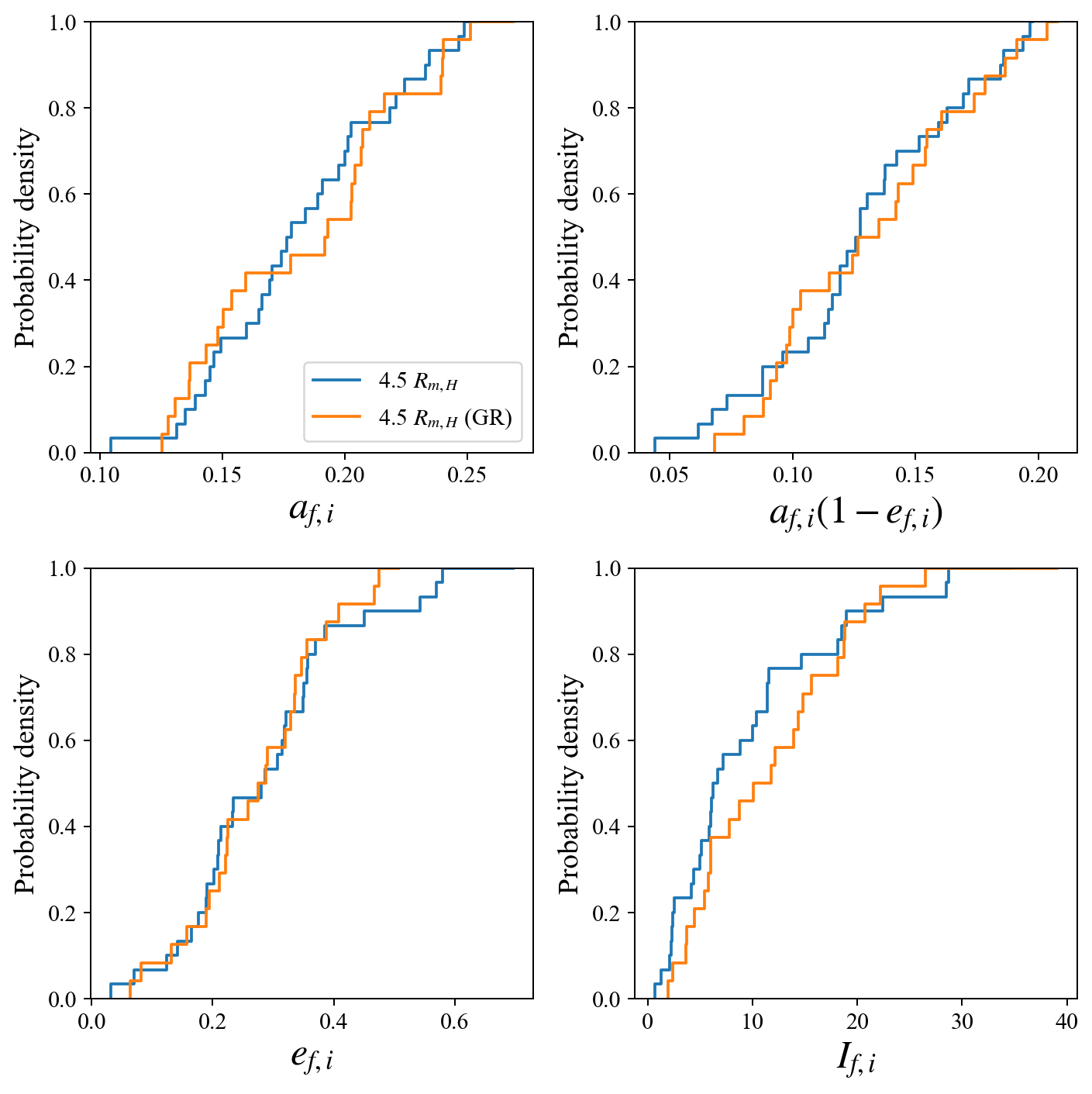}{0.45\textwidth}{(a)}
            \fig{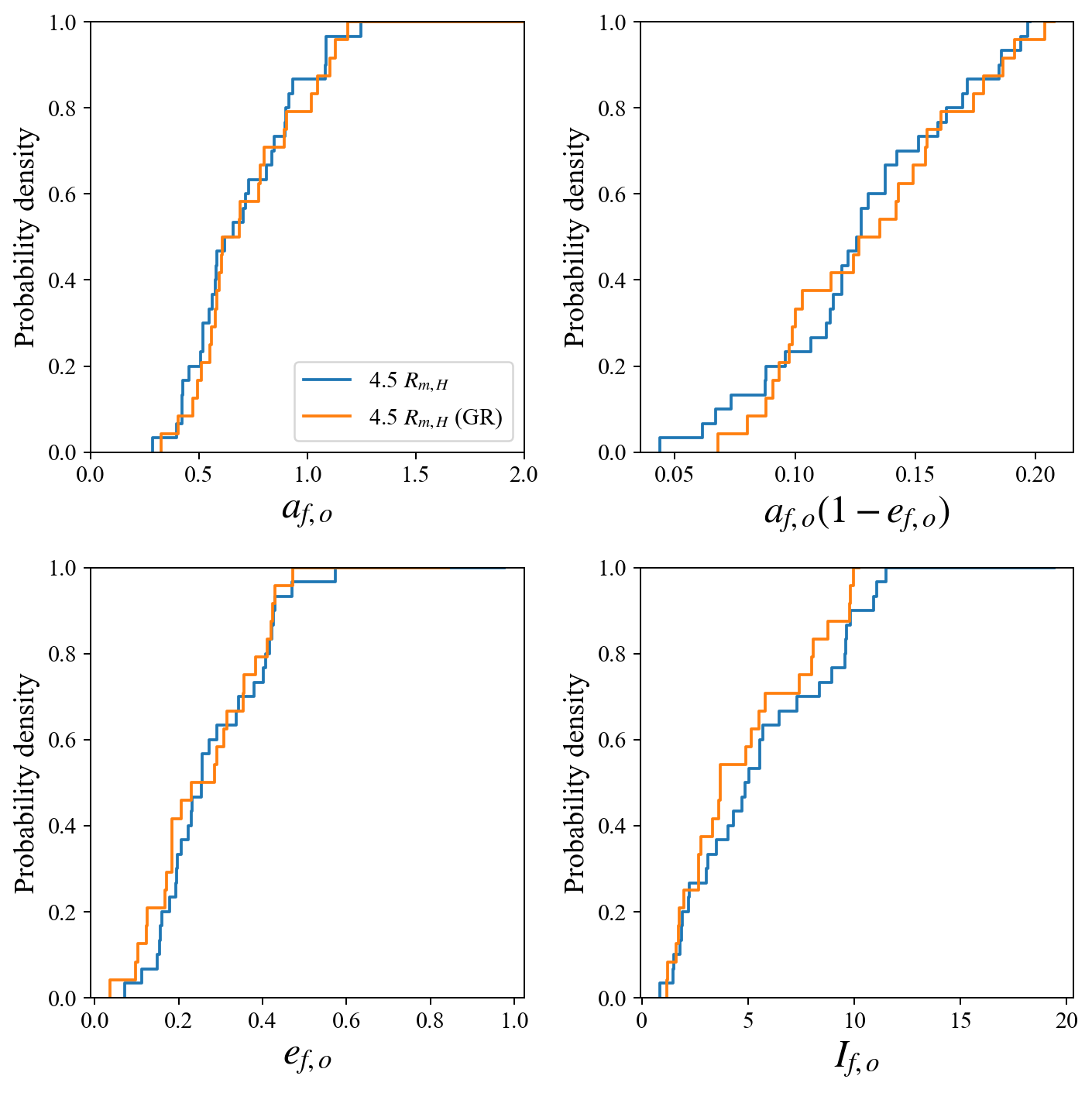}{0.45\textwidth}{(b)}
    }   
    \caption{Cumulative distributions of the semimajor axes, pericenter distances, eccentricities, and inclinations of the (\textit{left}) inner and (\textit{right}) outer planets for \textit{Sim.\ 2} of $K$4.5 (blue) and $K$4.5-GR (orange).}%
    \label{Figure/GR_vs_noGR_4.5RH}%
\end{figure*}

Figure \ref{Figure/GR_vs_noGR_4.5RH} (a) shows the cumulative distributions of the properties of the inner planet for the simulations $K$4.5 and $K$4.5-GR. The distributions of $a_{f,\,i}$ (upper left panel), $e_{f,\,i}$ (lower left panel), pericenter distance $a_{f,\,i} (1 - e_{f,\,i})$ (upper right panel), and inclination $I_{f,\,i}$ (lower right panel) are generally close to each other, but the inclination distribution for $K4.5$ is higher than that for $K4.5$-GR when $I_{f,\,i} \la 20^{\circ}$. Figure \ref{Figure/GR_vs_noGR_4.5RH} (b) shows the cumulative distributions of the orbital properties of the outer planet for the simulations $K$4.5 and $K$4.5-GR. The distributions are again close to each other. However, the lower right panel shows that the inclination distribution for $K4.5$ is lower than that for $K4.5$-GR when $I_{f,\,o} \ga 3^{\circ}$.

\begin{table*}[] 
\scriptsize
\tablewidth{0pt}
\caption{Two-sample Kolmogorov–Smirnov (KS) Test for the Cumulative Distributions of the Orbital Elements of the Inner and Outer Planet for $K$4.5 and $K4.5$-GR}
\centering
\label{t:KS_test_GR_VS_noGR}
\begin{tabular}{lcccccccc}
\noalign{\smallskip} \hline \hline \noalign{\smallskip}
 & $a_{f,\,i}$ & $e_{f,\,i}$& $I_{f,\,i}$ & $a_{f,\,i} (1- e_{f,\,i}$ ) & $a_{f,\,o}$ & $e_{f,\,o}$ & $I_{f,\,o}$ & $a_{f,\,o} (1- e_{f,\,o}$)\\
\hline
$D_{m,n}$ & 0.2219 & 0.1290 & 0.2619 & 0.1342 & 0.1316 & 0.1742 & 0.1652 &0.1626 \\
\noalign{\smallskip} \hline \noalign{\smallskip}
\end{tabular}
\tablecomments{When $\alpha$ is 0.05, the corresponding $D_{\text{crit}}$ is 0.365.}
\end{table*}

To quantitatively determine whether $K$4.5 and $K$4.5-GR produce significantly different results, we have performed the two-sample Kolmogorov-Smirnov (KS) test on the orbital elements of both cases. Suppose a sample has size $m$ with cumulative distribution $F(x)$, and the other sample has size $n$ with cumulative distribution $G(x)$. $D_{m,n}$ is the maximum difference between $F(x)$ and $G(x)$. The null hypothesis $H_{0}$ is that the two samples have the same distribution. $H_{0}$ should be rejected if $D_{m,n}$ $>$ $D_{\text{crit}}$, where $D_{\text{crit}}$ is the critical value defined as
\begin{equation} \label{D_crit}
 D_{\text{crit}} = c(\alpha) \sqrt{\frac{m+n}{mn}},
\end{equation}
where $\alpha$ is the significance level and $c(\alpha)$ is the inverse of the Kolmogorov distribution at $\alpha$. When we select $\alpha$ as 0.05, the corresponding $D_{\text{crit}}$ is 0.365. The calculated $D_{m,n}$ values for the orbital elements of the inner and outer planets are listed in Table \ref{t:KS_test_GR_VS_noGR}, in which we find no $D_{m,n}$ larger than $D_{\text{crit}}$. Thus, the KS test suggests no significant differences between $K$4.5 and $K$4.5-GR at the significance level of 0.05. 

\subsection{The influence of the mass of the hypothetical third planet}

We assume that the original GJ 1148 system has lost a hypothetical third planet via planet ejection in its history. Since the planet with the lowest mass is most likely the ejected one during close encounters \citep{chatterjee2008dynamical}, we assume that the mass of the third planet is not larger than the mass of the less-massive planet (GJ 1148 c with a mass of $0.227 M_{\text{J}}$) in the current GJ 1148 system. We have so far performed scattering experiments where we assume a third planet of mass $0.1 M_{\text{J}}$. Here we explore the situation where the lost planet has a mass of $0.227 M_{\text{J}}$, the same as GJ 1148 c. The other initial conditions and the integration schemes remain the same. Since the properties of the post-collision two-planet systems cannot account for the GJ 1148 system, we only analyze the post-ejection two-planet systems. We only fully integrate the post-ejection two-planet systems with the same mass order as the GJ 1148 system. To explore the influence of initial separation on scattering results, we perform three groups of simulations with $K$ = 3.5, 4.0, and 4.5, with 1400 simulations in each group. 

\begin{figure*}[!t]
    \centering
    \includegraphics[width=1\textwidth]{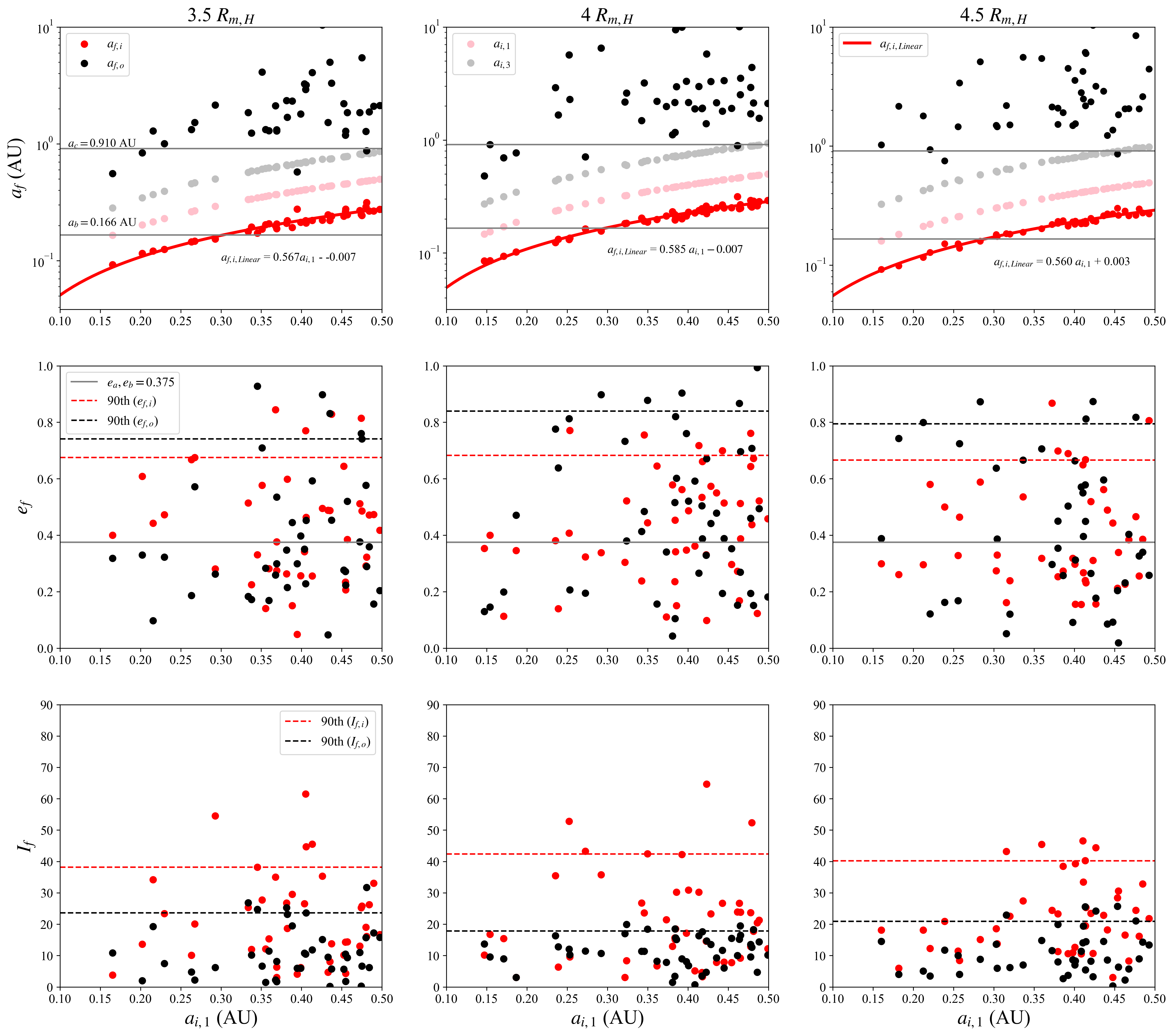}
    \caption{Properties of post-ejection two-planet systems with the ejected planet having a mass of $0.227 M_{\text{J}}$.}
    \label{fig:chap2:2P_EJ_difK_3rdp_0.227MJ}
\end{figure*}

Figure \ref{fig:chap2:2P_EJ_difK_3rdp_0.227MJ} shows the properties of two-planet systems arising from three-planet systems with the initial separation of 3.5, 4.0, and 4.5 $R_{m,\,H}$ by ejecting a $0.227 M_{\text{J}}$ planet. The upper panels show $a_{f,\, i}$ and $a_{f,\, o}$ versus $a_{i,\, 1}$. We can see that $a_{f,\, i}$ versus $a_{i,1}$ can be well-fitted by a linear relationship, and the slopes are similar for simulations with different $K$. However, the slopes are smaller than the simulations assuming a hypothetical $0.1 M_{\text{J}}$ planet in the original GJ 1148 system. This suggests that the inner planet can move to a smaller orbit if the ejected planet is more massive. Such a result is expected because the system's energy is conserved during the whole process \citep{marzari2002eccentric}. The total energy of the initial system is
\begin{equation} \label{tot_energy}
E = -\frac{GM_{\star}}{2}\left[\frac{m_{1}}{a_{1}} + \frac{m_{2}}{a_{2}} + \frac{m_{3}}{a_{3}}\right].
\end{equation}
If the hypothetical third planet is more massive, then the absolute value of total energy in the initial system is greater. Once the third planet is ejected, its orbital energy is negligible, and this energy is redistributed to the remaining two planets. The outer planet always moves to a distant orbit, and its energy gets smaller. Thus, the inner planet can move to a smaller orbit and dominate the energy of the final two-planet system.

The upper panels of Figure \ref{fig:chap2:2P_EJ_difK_3rdp_0.227MJ} show that systems with $a_{i,1}$ close to $0.29$ au can reproduce the observed semimajor axis ($0.166$ au) of the inner planet of GJ 1148, which is larger than $a_{i,1} \approx 0.21$ au in the simulations that lost a $0.1\,M_{\text{J}}$ planet. The semimajor axis of the outer planet in the final system also seems to increase as $a_{i,1}$ increases, but their range of values is extensive, similar to what we see in the $0.1\,M_{\text{J}}$ cases. In addition, we note that fewer systems have the outer planet's semimajor axis less than 0.910 au compared to the $0.1\,M_{\text{J}}$ cases, which indicates that the outer planet tends to scatter to a larger orbit when the ejected planet is more massive.

The distributions of eccentricities are shown in the middle panels of Figure \ref{fig:chap2:2P_EJ_difK_3rdp_0.227MJ}. We can see that the 90th percentiles of eccentricities of both planets are larger than 0.6 for all three groups of simulations. We also notice a more significant portion of systems with eccentricities larger than 0.375 compared to the $0.1\, M_{\text{J}}$ cases, suggesting that ejecting a more massive planet can increase the excitation of eccentricities. In the bottom panels, the 90th percentiles of the inclination of the inner planet are larger than those of the outer planet, similar to the $0.1\, M_{\text{J}}$ cases. In all three groups of simulations, the 90th percentiles of the inner planet are about $40^\circ$, while those of $0.1 \, M_{\text{J}}$ cases are less than $30^\circ$, suggesting that ejecting a more massive planet can also increase the excitation of inclinations. 

We did not study the scenario in which the hypothetical third planet has a mass greater than $0.227\, M_{\text{J}}$ for the following two reasons. First, less-massive planets are prone to being ejected during their dynamical evolution, implying that GJ 1148 c is likely to be ejected. Second, if a more massive planet were to be ejected, it would result in two-planet systems with even larger separations, further reducing the likelihood of finding a two-planet system similar to the GJ 1148 system.

\newpage
\subsection{Discrepancy with \cite{anderson2020situ}}

\cite{anderson2020situ} explored the origin of warm Jupiters via the \textit{in situ} planet-planet scattering scenario by using the $N$-body code REBOUND \citep{rein2012rebound}. They performed $N$-body simulations starting with a range of initial separations and planetary masses, including general relativistic (GR) apsidal precession. It is surprising that there are no post-ejection two-planet systems at all in their results. However, in our study, the search for GJ 1148-like planets is based on the existence of stable post-ejection two-planet systems. We have tried to check some results of the \textit{fiducial} $K = 4.0$ simulations in \cite{anderson2020situ} by repeating their simulations with the $N$-body codes SyMBA including the GR effect (SyMBA-GR) and REBOUND including the GR effect. The results of our simulations are presented in Appendix \ref{appendix:Reproduce_Anderson}. In particular, the simulations using the two $N$-body codes give similar results, and they produce a significant number (about $20\%$) of stable post-ejection two-planet systems. 

%%%%%%%%%%%%%%%% Conclusion %%%%%%%%%%%%%%%%%%%%%%%%%%%%%
\section{Conclusions} \label{sec:Conclusion}

We have explored a scenario in which the GJ 1148 system evolves from a three-planet system. Dynamical instabilities can lead to close encounters between the planets, resulting in planet-planet collisions, planet-star close approaches, or planet ejections. We first assumed that there is a third planet with a mass of 0.1 $M_{\text{J}}$ in the original GJ 1148 system. By performing scattering experiments with initial orbital separations of 3.5, 4, and 4.5 $R_{m, H}$, we found that most scattering outcomes are planet-planet collisions, followed by planet ejections, and then a planet getting too close to the star. The properties of the resulting two-planet systems show that the post-collision two-planet systems are unlikely to reproduce the GJ 1148 system because they have small orbital spacings and small eccentricities. However, the post-ejection two-planet systems have more similar properties to the GJ 1148 system. 

The semimajor axis of the final inner planet has a strong linear relationship with the semimajor axis of the initial innermost planet. We found that systems with $a_{i,1}$ close to 0.21 au can reproduce the semimajor axis of the inner planet of GJ 1148. However, the outer planet's semimajor axis varies in a wide range and requires more simulations to search for ones that can match well the observed system. By performing simulations with $K = 4.5$ in a narrower range of the semimajor axis of the initial innermost planet, we found several systems that are similar to GJ 1148, including one that can reproduce the best-fit properties of the GJ 1148 planets (but with both planets on slightly larger orbits than the GJ 1148 planets).

We also checked the influence of GR precession on ejection outcomes. The Kolmogorov–Smirnov (KS) test shows that GR precession does not significantly affect the properties of post-ejection two-planet systems. To explore the influence of the mass of the ejected planet, we performed simulations with a hypothetical third planet having the same mass as GJ 1148 c (0.227 $M_{\text{J}}$). We found that the optimal initial $a_{i,1}$ moves to around 0.29 au, and the excitation of eccentricities and inclinations is increased.

\begin{acknowledgments}
We thank Trifon Trifonov and Dong Lai for their informative discussions. This work was supported by a postgraduate studentship at the University of Hong Kong (L.Y.) and Hong Kong RGC grant 17306720 (L.Y. and M.H.L.). L.Y. is grateful for a grant support from the ETH Research Commission, no ETH-05 22–1.

\end{acknowledgments}

\appendix
\section{Simulations of \textit{In situ} scattering of warm Jupiters}  \label{appendix:Reproduce_Anderson}

In this Appendix, we introduce the simulations designed to reproduce part of the results in the \textit{fiducial} group of \cite{anderson2020situ}. We perform two sets of simulations, consisting of 300 simulations each, using SyMBA-GR and REBOUND, respectively. We use the same parameters and integration strategy as the \textit{fiducial} group of \cite{anderson2020situ}. The initial planetary system has three planets with masses 0.5, 1.0, and 2.0 $M_{\text{J}}$, orbiting around a star with the same mass and radius as the Sun. For each planet, we sample the initial eccentricity in [0.01, 0.05], inclination in the range of [$0^{\circ}$, $2^{\circ}$], and the orbital phase angles uniformly in [$0^{\circ}$, $360^{\circ}$]. The planets are placed with random ordering and spacing of $K =  4.0$. For each planetary system, we first integrate it for a time span of $10^{6}\, P_{i}$ (\textit{Phase 1}), and then continue to integrate it for a time span of $10^{8}\, P_{in}$ (\textit{Phase 2}).
For the simulations using REBOUND, we follow \cite{anderson2020situ} and use the IAS15 integrator for \textit{Phase 1} and the Mercurius integrator for \textit{Phase 2}.

\begin{figure*}[!t]
    \centering
    \includegraphics[width=1\textwidth]{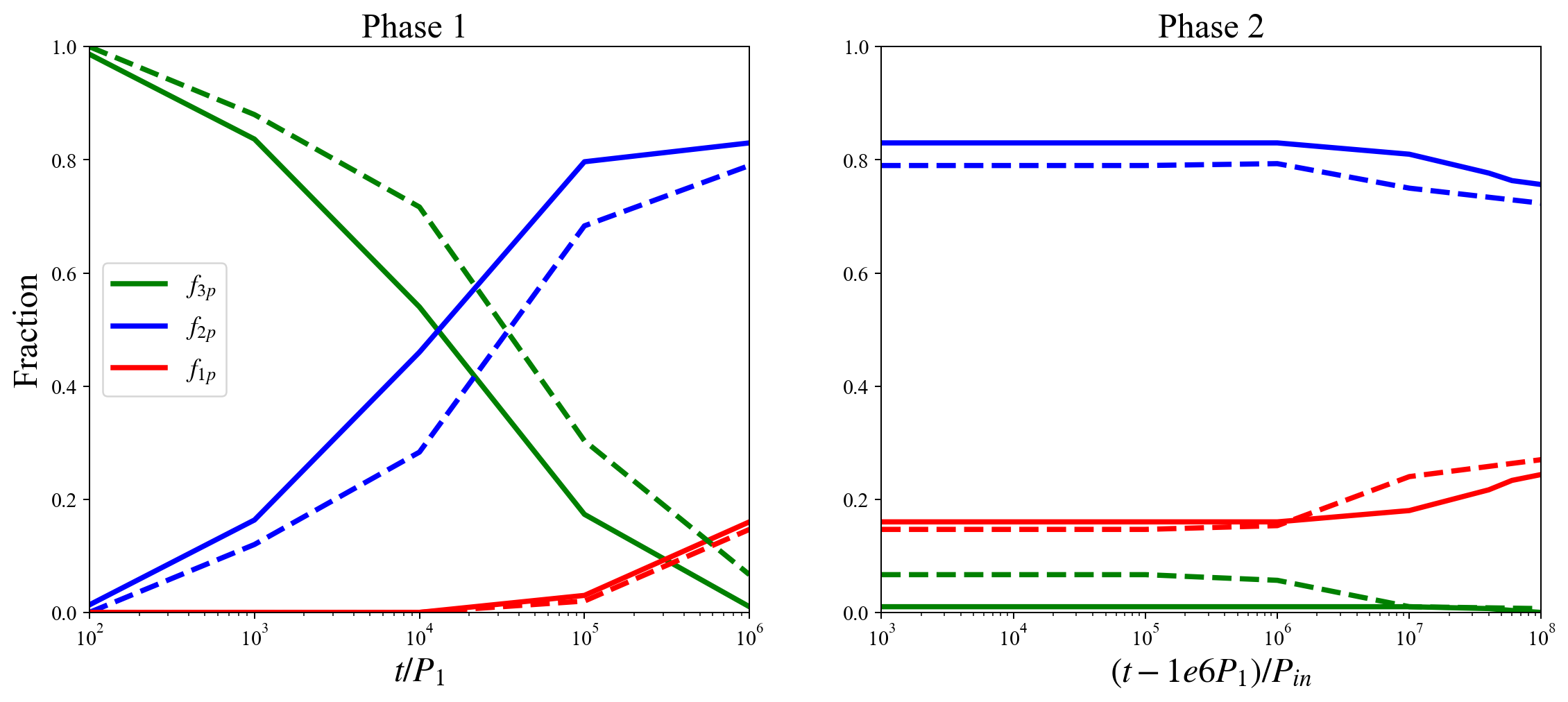}
    \caption{The fractions of one-, two-, and three-planet systems as a function of time for the \textit{fiducial} simulations using SyMBA-GR (solid lines) and REBOUND (dash lines). For comparison, it is plotted to look similar to Figure 1 of \cite{anderson2020situ}. \textit{Left}: \textit{Phase 1}, in which we integrate the initial  three-planet systems for a time span of $10^6$ periods of the initial inner planet. \textit{Right}: \textit{Phase 2}, in which the remaining two- and three-planet systems are integrated for a time span of $10^8$ orbital periods of the inner planet.}    \label{fig:Fraction_symba_vs_rebound}
\end{figure*}

Figure \ref{fig:Fraction_symba_vs_rebound} shows the time evolution of the fractions of one-, two-, and three-planet systems integrated by both SyMBA-GR and REBOUND. For both sets of simulations, nearly all three-planet systems become unstable at the end of \textit{Phase 1}. The fractions of  two-planet  and one-planet systems increase to around $80\%$ and $20\%$, respectively. These fractions shown in the diagram of \textit{Phase 1} are comparable to the left panel in Figure 1 of \cite{anderson2020situ}. However, at the end of \textit{Phase 2}, the fraction of two-planet systems decreases to 75\% for simulations with SyMBA-GR (72\% for simulations with REBOUND), while only 48\% two-planet systems remain stable as shown in  Figure 1 of \cite{anderson2020situ}. We find that the striking discrepancy in the final fraction of two-planet systems is mainly attributed to the existence of a nontrivial fraction ($\sim 20\%$) of stable post-ejection two-planet systems (see $N_{ej,2p}$ in Table \ref{t:ScaterOutcomesReproduceAnderson}) in our simulations, while there is no such system in the \textit{fiducial} group of \cite{anderson2020situ}.
This discrepancy of $\sim 20\%$ is clearly significant, even though our number of simulations ($300 + 300$) is much smaller than $\sim 3300$ of \cite{anderson2020situ}.
We have also checked that the vast majority of the post-ejection two-planet systems are stable on even longer timescales.
When we extend our \textit{Phase 2} integrations using SyMBA-GR 10 times longer to $10^9\, P_{in}$, $N_{ej,2p}$ decreases only slightly from $56$ in Table \ref{t:ScaterOutcomesReproduceAnderson} to $51$.

\begin{table*}[htb]
\scriptsize
\tablewidth{0pt}
\centering
\caption{Scattering Outcomes of Simulations Starting with the Same Parameters as the \textit{Fiducial} Group of \cite{anderson2020situ}   }
\label{t:ScaterOutcomesReproduceAnderson}
\begin{tabular}{lccccccccccccc}
\hline \hline
Integrator & $N_{\text{total}}$ & $N_{pp,2p}$ & $N_{ps,2p}$& $N_{ej,2p}$ & $N_{pp,1p}^{(1)}$ & $N_{ps,1p}^{(1)}$ & $N_{ej,1p}^{(1)}$ & $N_{pp,1p}^{(2)}$ & $N_{ps,1p}^{(2)}$ & $N_{ej,1p}^{(2)}$ & $N_{F,\,1P}$ & $N_{F,\, 2P}$ & $N_{F,\,3P}$\\

\hline
SyMBA-GR & 300 & 163 & 5  &56  & 53 & 11 & 12 
& 3 & 13 & 60  & 76 & 224 & 0 \\
REBOUND & 300  & 151 & 0  &64  & 48 & 4  & 33 
& 36 & 7 & 42  & 85 & 215 & 0 \\
\hline
\end{tabular}
\end{table*}

\bibliography{Reference2}{}
\bibliographystyle{aasjournal}

\end{document}